\documentclass[12pt,a4paper]{article}

\usepackage{axodraw4j}
\usepackage{pstricks}
\usepackage{color}
\usepackage{epsfig}

\begin{document}
\def \be  {\begin{equation}}
\def \ee  {\end{equation}}
\def \beq  {\begin{equation}}
\def \eeq  {\end{equation}}
\def \ba  {\begin{eqnarray}}
\def \ea  {\end{eqnarray}}
\def \baa {\begin{eqnarray*}}
\def \eaa {\end{eqnarray*}}
\def \lab #1 {\label{#1}}
\newcommand\bqa {\begin{eqnarray}}
\newcommand\eqa {\end{eqnarray}}
\newcommand\pr {\partial}
\newcommand\apr {\overline {\partial }}
\newcommand\nn {\nonumber}
\newcommand \noi {\noindent}
\newcommand{\bear}{\begin{array}}
\newcommand{\enar}{\end{array}}
\newcommand{\hf}{\frac{1}{2}}
\newcommand{\vx}{\vec{x}}
\newcommand{\R}{\mathbb{R}}
\newcommand{\C}{\mathbb{C}}
\newcommand{\Q}{\mathbb{Q}}
\newcommand{\F}{\mathbb{F}}
\newcommand{\A}{\overline{\mathbb{C}}}
\newcommand{\Z}{\mathbb{Z}}
\newcommand{\bg}{{\bf g}}
\newcommand{\Tpl}{{T}_+}
\newcommand{\Tmin}{\mathcal{T}_-}
\newcommand{\LL}{{L}}
\newcommand{\AAA}{\overline{A}}
\newcommand{\inv}[1]{{#1}^{-1}} 

\def\t{\theta}
\def\T{\Theta}
\def\w{\omega}
\def\ov{\overline}
\def\a{\alpha}
\def\b{\beta}
\def\g{\gamma}
\def\s{\sigma}
\def\l{\lambda}
\def\wt{\widetilde}

\def \CO {{\cal O}}
\def \CP {{\cal P}}
\def \CT {{\cal T}}
\def \CM {{\cal M}}
\def \CK {{\cal K}}
\def \CH {{\cal H}}
\def \CI {{\cal I}}
\def \CV {{\cal V}}
\def \CJ {{\cal J}}
\def \CL {{\cal L}}

\font\cmss=cmss12 \font\cmsss=cmss10 at 11pt
\def\inbar{\,\vrule height1.5ex width.4pt depth0pt}
\def\IC{\relax\hbox{$\inbar\kern-.3em{\rm C}$}}
\def\IZ{\relax{\hbox{\cmss Z\kern-.4em Z}}}
\def\IR{{\hbox{{\rm I}\kern-.2em\hbox{\rm R}}}}
\def\R{{\tiny \IR}}
\def\IP{{\hbox{{\rm I}\kern-.2em\hbox{\rm P}}}}
\def\II{\hbox{{1}\kern-.25em\hbox{l}}}

\begin{titlepage}

\hfill\parbox{40mm}
{\begin{flushleft}  ITEP-TH-14/09\\
FTPI-MINN-09/11
\end{flushleft}}

\vspace{10mm}

\centerline{\large \bf One-loop derivation of the}
\centerline{\large \bf Wilson polygon - MHV amplitude duality}

\vspace{17mm}

\centerline{\bf A. Gorsky$^{1,4}$ and A. Zhiboedov$^{1,2,3}$}
\vspace{10mm}

\begin{center}
{\it $^1$ Institute of Theoretical and Experimental Physics, Moscow, Russia, \\
$^2$ Joint Institute for Nuclear Research, Bogoliubov Laboratory of Theoretical Physics, Dubna, Russia, \\
$^3$ Moscow State University, Physics Department, Moscow, Russia, \\
$^4$ FTPI, University of Minnesota.} \\

\vspace{1cm}
\end{center}

\vspace{1cm}

\centerline{\bf Abstract}
We discuss the origin of the Wilson polygon - MHV amplitude duality
at the perturbative level. It is shown that the
duality for the MHV amplitudes at  one-loop level can be proven upon the peculiar
change of  variables in
Feynman parametrization and the use of the relation between  Feynman integrals
at the different
space-time dimensions. Some generalization of the duality which implies the
insertion of the particular vertex operator at the Wilson triangle is found
for the 3-point function. We discuss analytical structure of Wilson loop diagrams
and present the corresponding Landau equations.
The geometrical interpretation of the loop diagram
in terms of the hyperbolic geometry is discussed.


\end{titlepage}

\section{Introduction}

The clarification of the geometrical structure behind the perturbation theory in SYM which would provide the way of summation of
the series remains the challenging problem.  During the last years two novel ideas
concerning these issues have been developed. It was demonstrated in
\cite{witten} that the important localization phenomena happens
for the perturbative amplitude in the twistor space. On the other hand
the stringy calculation of the amplitudes \cite{am} suggested the hidden
duality between the amplitudes in ${\cal N}=4$ SYM and the Wilson polygon built from the light-like momenta of the
external gluons. It is important to note to that amplitudes look like to be mapped to ordinary position space Wilson loop.
A connection between amplitudes and momentum space Wilson loops was investigated in \cite{pm}.

This duality has been checked
at one \cite{kor,trav} and two loops \cite{six1,six2} in the perturbative theory  and  has the chance to be all-loop exact (see \cite{alday} for the review).
During this development it was also realized
that the important dual superconformal symmetry is under the carpet
which was clarified both in the weak coupling \cite{kor2,kor3} and strong coupling sides \cite{kom} of the correspondence.
The dual superconformal symmetry was argued to be the consequence of the fermionic T-duality in the stringy sigma
model \cite{ber} and the combination of the usual superconformal and dual superconformal symmetries
implies the Yangian symmetry in the perturbative
${\cal N}=4$ SYM theory \cite{yangian}.

In spite of the impressive progress many key issues are still to be clarified.
In this paper we shall focus on the origin of the Wilson polygon- MHV amplitude
duality which shall be analyzed at the one-loop level. We shall try to get
the precise mapping between the  one-loop diagram for MHV amplitude and the one-loop
correction to the Wilson polygon. It turns out that upon the proper
change of variables in the Feynman parametrization of the loop integral
for the six-dimensional box diagram brings it to the form of the Wilson polygon in the
four dimensions. Oppositely the four dimensional box diagram
can be related with the Wilson polygon is six dimensions.
The IR divergences of the amplitudes get mapped into the
UV divergences of the Wilson polygon. Moreover, it is seen that the MHV amplitude
obeys this special property since
it can be expressed in terms of the two mass-easy box diagrams only and  simple
change of variables we have found does not work
for the non-MHV amplitudes. Using known interplay between
particular $D=6$ and $D=4$ integrals \cite{Tarasov, Nizic2} the answer can be immediately presented
in terms of the finite part of the $D=4$ two-mass easy box.

The  loop amplitudes can be calculated via dispersion relations hence
the duality implies that some version of the imaginary part calculations can be
formulated for the loop corrections to the Wilson polygon as well. To this aim
we shall slightly generalize the cut technique for the loop diagrams and shall argue
that on the Wilson polygon side the dispersion calculation corresponds to the
cutting of the Wilson polygon into the several pieces and the following gluing with the
insertion of particular operators. We shall also comment on the Landau equations
for the singularities on the Wilson polygon side.

It is natural to search for more natural geometry behind the one-loop calculation
which would shed additional light on the duality under discussion. Let us first
comment on the previous studies of this issue.
The one-loop correlation functions suggest
the natural emergence of the $AdS$ type geometry in three-point \cite{gopa}
and four-point functions \cite{aha}. Similar hyperbolic structure is also clearly seen in the one-loop
effective action in the constant external field
\cite{gorly}.  In both cases the Schwinger parametrization of the loop integral plays the crucial role.
In particular for the three-point function the combination of the Schwinger parameters plays
the role of the radial coordinate in the $AdS_5$  \cite{gopa} while in the effective action case
similar identification  emerges in the $AdS_3$ submanifold \cite{gorly}.

The geometry behind
the BDS formula \cite{bds} emerging upon summation over the loops has been suggested in \cite{gor}
and the corresponding fermionic representation which supports the hidden integrability has been found.
The key point is that there is natural playground for the topological strings both in the $A$ model
with the Kahler gravity and $B$ model involving KS gravity on the moduli space
of the complex structures. The both complex and K\"{a}hler
types of moduli are provided by the kinematical invariants of the scattering particles.

In this paper we shall mention  geometrical aspect of the one loop calculation. based on the
observation of \cite{dav} related with the Kahler moduli side.
It was found in \cite{dav} that the one loop box integral counts the
hyperbolic volume of the 3d manifold in the space of Feynman parameters. Contrary to the Gopakumar's
approach when the four-point function is treated differently from the three-point function in
this approach they are considered on the equal footing. Since the 3d hyperbolic manifolds emerge
naturally as the knot complements we shall make some links with the Chern-Simons calculation
with the inserted Wilson loop.

The paper is organized as follows. In Section 2 we review the duality between the MHV amplitudes
in ${\cal N}=4$ theory and the Wilson polygon. In Section 3 we briefly explain the relevant hyperbolic geometry
behind the one-loop calculations. Section 4 is devoted to the
explicit derivation of the duality for MHV amplitude at the one loop level.
In Section 5 we provide the simplified example
of the duality for the 3-point function which involves the vertex operator on the Wilson polygon side.
In Sections 6  we consider some aspects of the unitarity calculation of the Wilson polygons.
Section 7 is devoted to the comments concerning the relation of the divergent
contributions with the hyperbolic geometry
of one-loop diagrams.
In the last Section we shall collect our observations and mention the open problems.

\section{The connection between Wilson polygons and MHV amplitudes}

In this section we briefly review the conjectured duality
between the loop amplitudes in ${\cal N}=4$ theories and Wilson polygons built
from the external momenta (see \cite{alday} for review).

Precisely, it was conjectured in \cite{am} that any MHV  $N$-leg amplitude
follows from the vacuum expectation value of the Wilson loop of the
special form
\beq
\frac{A^{MHV}_{all-loop}}{A^{MHV}_{tree}} =<W(p_1,p_2,...,p_N)>
\label{am}
\eeq
where the closed Wilson loop polygon has light-like
momenta at the edges and vertices at $x_i$. Its closeness is provided
by the total momentum conservation. At the strong coupling limit
both Wilson polygon as well as the MHV amplitude are calculated in the
sigma model approach.

At weak coupling to check this polygon- MHV amplitude duality one
considers the expansion of the Wilson polygon in the YM coupling
treating Wilson loop as placed in the coordinate space. In its weaker
form the duality takes form
\beq
Fin[\frac{A^{MHV}_{all-loop}}{A^{MHV}_{tree}}]=Fin[<W(p_1,p_2,...,p_N)>]
\label{am}
\eeq
The perfect matching of the Wilson loop and amplitudes has been found for one \cite {kor,trav} and
two loop answers up to six external legs \cite{six1,six2}. Moreover, it
was demonstrated that the anomalous Ward identities for the special
conformal transformations of the form

\beq K^{\nu}W(x_1,\dots x_N)= \sum_{i=1}^{n}(2x_i^{\nu}x_i \partial_i -x_i^2 \partial_i^{\nu}) W(x_1,\dots x_N)          =
\frac{1}{2}\Gamma_{cusp}\sum_{i=1}^n ln \frac{x_{i,i+2}^2}{x_{i-1,i+1}^2} x_{i,i+1}^{\nu}
\eeq
where $\Gamma_{cusp}$ is the cusp anomalous dimension,
fix the answer up to four external legs .

The anomalous
Ward identities  can be applied both to the amplitudes and the Wilson polygons, however,
starting with six external legs the Ward identity allows the arbitrary
function of the conformal ratios, which can not be fixed by the superconformal
group arguments.

There is some specifics concerning the loop MHV amplitudes.
The one-loop Wilson loop diagram with the arbitrary number of external legs
can be mapped to the finite part of two-mass easy box which is the main building block
of the answer. The generalization of the duality to the non-MHV
amplitudes  turns out to be nontrivial issue. In particular it is known \cite{bernNMHV} that
the NMHV loop amplitude involves 3 mass box diagrams as well and
harder diagrams are relevant for the $N^k MHV$ amplitudes. No recipe
for the duality beyond the MHV case has been formulated yet.

It was demonstrated that the unitarity approach is fruitful for the
description of the loop amplitudes. General planar color-ordered
one-loop scattering super-amplitude can be written in the following way

\beq
{\cal A}_{n;1} =i (2 \pi)^4 \delta^{4}(p) \sum ( {\cal C}^{4m} I^{4m} + {\cal C}^{3m} I^{3m} + {\cal C}^{2mh} I^{2mh}+{\cal C}^{2me} I^{2me}+{\cal C}^{1m} I^{1m})
\eeq
where $I$'s are the scalar-box integrals with the corresponding number
of legs being off-shell.

The only thing one needs to calculate for given amplitude is the coefficients,
which can be done in terms of quadruple cuts.
The general form of the ${\cal C}^{m}$ takes the form

\beq
{\cal C}^{m}= \delta^{8} (\sum_{i=1}^{n} \lambda_{i} \eta_{i}) [{\cal P}^{(0),m}_{n;1}+ {\cal P}^{(4),m}_{n;1}+...+{\cal P}^{(4 n - 16),m}_{n;1}]
\eeq
where ${\cal P}^{(4 k),m}_{n;1}$'s are homogenous polynomials of degree $4 k$ in Grassmann variables.

One-loop MHV super-amplitude takes the following form

\beq
{\cal A}^{MHV}_{n;1} =i (2 \pi)^4 \delta^{4}(p) \frac{\delta^{8} (\sum_{i=1}^{n} \lambda_{i} \eta_{i})}{\langle 12 \rangle \langle23\rangle...\langle n1\rangle} [ \sum_{s=3}^{n-1} I^{2me}_{1,2,s,s+1} \Delta_{1,2,s,s+1} + cyclic]
\eeq
where $\Delta_{r,t,s,s+1}=- \frac{1}{2} [ x^{2}_{s r} x^{2}_{s+1  t} - x^{2}_{s+1  r} x^{2}_{s t}]$. The answer is fully defined by
two-mass easy boxes.

The general one-loop NMHV amplitude has  more complicated structure, namely

\beq
{\cal A}^{NMHV}_{n;1} = {\cal A}^{MHV}_{n;0}[ \sum_{p,q,r=1}^{n} R_{pqr} (1 + \frac{\lambda}{8 \pi^2} V_{pqr} + O(\varepsilon))]
\eeq
where two-mass hard and three-mass boxes are involved, $R_{pqr}$ are dual superconformal and $V_{pqr}$ are dual conformal invariants.

\section{Hyperbolic geometry of one loop}

In what follows it will be useful to utilize the geometrical
picture behind the one-loop calculations which we shall
review  following \cite{dav}.
Let us explain first the explicit map of the box diagram to the
hyperbolic volume of the particular simplex build from the kinematical
invariants of the external momenta. To this aim introduce
the Feynman parametrization of the internal generically massive propagators
with the parameters $\alpha_i$.  If one considers the one-loop N-point function
with the external momenta $p_i$
in D space-time dimensions it can be brought into the usual form
\beq
J(D,p_1,\dots p_N)\propto
\int_{0}^{1} \dots \int_{0}^{1}\prod d\alpha_i \delta(\sum \alpha_i -1)
[\sum \alpha_i^2 m_i^2 +\sum_{j<l} \alpha_i \alpha_j m_j m_l C_{il}]^{D/2-N}
\eeq
where
\beq
C_{jl}=\frac{m_i^2 +m_l^2 -k_{jl}^2}{2m_j m_i},\qquad k_{ij}= p_i -p_j
\eeq
and $m_i$ is the mass in the i-th propagator.

It is possible \cite{dav}  to organize for the generic
one-loop diagram the $N$ dimensional simplex defined as follows.
First introduce  the $N$ mass vectors $m_i a_i$ , where $a_i$ are the unit vectors. The
length of the side connecting the i-th and j-th mass vectors is $\sqrt {k_{ij}}$
that is one can define the momentum side of the simplex. Therefore the $N$-dimensional simplex involves
$\frac{N(N+1)}{2}$ sides including
N mass sides as well as $\frac{N(N-1)}{2}$ momentum sides. At each vertex $N$ sides meet and at all vertices
but one there are one mass side and $(N-1)$ momentum sides. The volume of such $N$-dimensional simplex
is given as follows
\beq
V^{(N)}=\frac{(\prod m_i)\sqrt {det C}}{N!}
\eeq
There are $(N+1)$ hypersurfaces of dimension $(N-1)$ one of which contains only momentum sides
and can be related with the massless N-point function.

It is convenient to make the change of variables which transforms the loop integral
into the following form
\beq
J (D,p_1,\dots p_N)\propto \prod m_i^{-1}
\int_{0}^{\infty} \dots \int_{0}^{\infty}\prod d\alpha_i \delta(\alpha^{T} C \alpha -1)
(\sum \frac{\alpha_i}{m_i})^{N-D}
\eeq
that is integration now is over the quadrics in the space of the Feynman parameters.
It is useful to introduce the content of the $N$-dimensional solid angle $\Omega^{(N)}$
subtended by the hypersurfaces at the mass meeting point. It turns out that $\Omega^{(N)}$
coincides with the content of the $(N-1)$ dimensional simplex in the hyperbolic space
whose sides are equal to the hyperbolic angles $\tau_{ij}$ defined at small masses as follows
\beq
C_{ij}=cosh\tau_{ij}
\eeq
Then the integral for the case $D=N$ acquires the following form
\beq
J(N,p_1,\dots p_N)=i^{1-2N}\frac{\pi^{N/2}\Gamma(N/2) \Omega^{(N)}}{N! V^{(N)}}
\eeq
hence the calculation of the Feynman integral is nothing but the
calculation of the hyperbolic volume in the proper space. The case
$N\neq D$ can be treated similarly with some modification \cite{dav}.

Let us turn now to the case of interest that is N-leg MHV amplitudes
in four dimensions. The crucial point is that one-loop MHV amplitudes
can be presented as the sum of the two mass-easy box diagrams. This
diagrams are IR divergent that is it is useful to start with
the box diagram with all off-shell particles.
We have the situation with $D=N$ simplices in the hyperbolic space.
\beq
J(4,p_1,p_2,p_3,p_4)= \frac{2i\pi ^2 \Omega^{(4)}}{m_1 m_2 m_3 m_4 \sqrt{det C}}
\eeq
and since  all internal propagators are massless in our case
we get the ideal hyperbolic tetrahedron whose all vertices are
at infinity. In the massless limit we get
\beq
(m_i^2 m_2^2 m_3^2 m_4^2 det C)_{m_i\rightarrow 0}=
\frac{1}{16} \lambda(k_{12}^2 k_{34}^2, k_{13}^2 k_{24}^2, k_{14}^2 k_{23}^2)
\eeq
where the K\"{a}llen function $\lambda(x,y,z)$ is defined as
\beq
\lambda(x,y,z)=x^2 +y^2 +z^2 -2xy -2yz -2zx
\eeq
and $\sqrt{-\lambda}$  is just the area of the triangle with sides
$\sqrt{k_{12}^2k_{34}^2}, \sqrt{k_{23}^2k_{24}^2}, \sqrt{k_{31}^2k_{23}^2}$. The hyperbolic volume
of the ideal tetrahedron under consideration reads as
\beq
2i\Omega^{(4)}= Cl_2(\psi_{12}) +Cl_2(\psi_{13}) +Cl_{2}(\psi_{23})
\eeq
where the dihedral angles are defined via the kinematical invariants
\beq
-cos \psi_{12}= \frac{k_{13}^2 k_{24}^2 + k_{14}^2 k_{23}^2 - k_{12}^2 k_{34}^2 }
{\sqrt{k_{13}^2 k_{23}^2 k_{14}^2 k_{43}^2 }}
\eeq
\beq
-cos \psi_{13}= \frac{k_{14}^2 k_{23}^2 + k_{12}^2 k_{43}^2 - k_{13}^2 k_{24}^2 }
{\sqrt{k_{14}^2 k_{23}^2 k_{12}^2 k_{43}^2 }}
\eeq
\beq
-cos \psi_{14}= \frac{k_{12}^2 k_{34}^2 + k_{13}^2 k_{24}^2 - k_{14}^2 k_{32}^2 }
{\sqrt{k_{13}^2 k_{24}^2 k_{12}^2 k_{43}^2 }}
\eeq
and $\psi_{12}=\psi_{34},\quad \psi_{13}=\psi_{24},\quad \psi_{14}=\psi_{32}$.
The functions involved are defined as
\beq
Cl_2(x)=Im [Li_2(e^{ix})= - \int_{0}^{x} dy ln|2sin y/2|
\eeq
In the case of the two mass-easy box diagram defining the one-loop
MHV amplitude the additional simplification of the kinematical invariants
happens since two external particles are on the mass shell. In this case
the arguments of the $Li_2$ function degenerates to the conformal
ratios of four points. The geometrical picture behind the divergent part
of the diagram will be discussed later.

Note that massless four-point box answer coincides with the three-point result
which is known for a while \cite{dav2}. However the geometrical object
responsible for the three-point function is just the triangle. The answer for the generic
three-point function is expressed in terms of the angles of the basic triangle only \cite{dav}.

The appearance of the hyperbolic volume implies that the topological string approach
or CS with $SL(2,C)$ group are relevant \cite{hikami}. Indeed we can consider the ideal tetrahedron
as the knot complement and shall calculate it via the Chern-Simons theory action
with the complex group. It turns out that the choice of the particular
values of the kinematical invariants corresponds to the choice of
particular knot \cite{brod}.

\section{Derivation of the Wilson polygon - MHV amplitude duality at one loop}
In this Section we shall derive the duality at one-loop lever via the two step procedure.
First we describe the change of variables in the space of Feynman parameters
which brings the two mass-easy box diagrams into the form of the Wilson polygon in
the different dimension. Than we  take use of the relation between the
Feynman diagrams in $D=6$ and $D=4$.

Let us start with the definition of general box in $D_{IR}= d_{IR} - 2 \epsilon_{IR}$ dimensions.

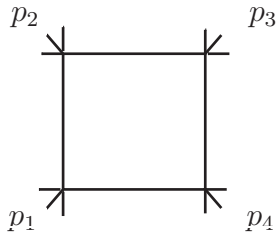
\begin{center}
\fcolorbox{white}{white}{
  \begin{picture}(122,93) (92,-32)
    \SetWidth{1.0}
    \SetColor{Black}
    \CBox(109,-21)(162,30){Black}{White}
    \Text(89,-37)[lb]{\normalsize{\Black{$p_{1}$}}}
    \Text(90,40)[lb]{\normalsize{\Black{$p_{2}$}}}
    \Text(179,40)[lb]{\normalsize{\Black{$p_{3}$}}}
    \Text(178,-37)[lb]{\normalsize{\Black{$p_{4}$}}}
    \Line(101,30)(109,30)
    \Line(162,-21)(170,-21)
    \Line(100,-21)(108,-21)
    \Line(109,40)(109,31)
    \Line(162,39)(162,30)
    \Line(163,30)(171,30)
    \Line(162,-21)(162,-30)
    \Line(109,-22)(109,-31)
    \Line(103,37)(109,30)
    \Line(162,-21)(168,-28)
    \Line(162,30)(168,37)
    \Line(103,-28)(109,-21)
  \end{picture}
}
\end{center}
and use notations from \cite{Nizic}
\begin{eqnarray}
I(p_{i},D_{IR},\mu_{IR}) &=& - i \pi^{-\frac{D_{IR}}{2}} (\mu_{IR}^2)^{\epsilon_{IR}} \int d^{D_{IR}}l \frac{1}{l^2 (l - p_1)^2 (l-p_1-p_2)^2 (l+p_4)^2} \\ \nonumber
p_{i}^2 &=& m_{i}^2 \nonumber
\end{eqnarray}
One can introduce Feynman parameters and take the integral over $l$ which amounts to

\begin{eqnarray}
I(p_{i},D_{IR},\mu_{IR}) &=& (\mu_{IR}^2)^{\epsilon_{IR}} \Gamma(4 - \frac{D_{IR}}{2}) \int \prod d x_{i} \frac{\delta( 1 - x_1 - x_2 - x_3 - x_4)}{(-\Delta)^{4 - \frac{D_{IR}}{2}}} \\ \nonumber
\Delta &=& s x_1 x_3+ u x_2 x_4 +m_{1}^2 x_{1} x_{2} + m_{2}^2 x_{2} x_{3} + m_{3}^2 x_{3} x_{4}+ m_{4}^2 x_{4} x_{1} \\ \nonumber
\end{eqnarray}
where $s=(p_1+p_2)^2$ and $u=(p_2+p_3)^2$.


Let us focus on the two-mass easy box diagram

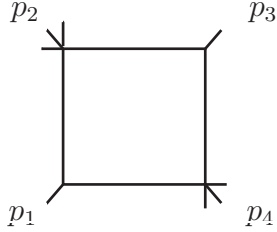
\begin{center}
\fcolorbox{white}{white}{
  \begin{picture}(122,93) (92,-32)
    \SetWidth{1.0}
    \SetColor{Black}
    \CBox(109,-21)(162,30){Black}{White}
    \Text(89,-37)[lb]{\normalsize{\Black{$p_{1}$}}}
    \Text(90,40)[lb]{\normalsize{\Black{$p_{2}$}}}
    \Text(179,40)[lb]{\normalsize{\Black{$p_{3}$}}}
    \Text(178,-37)[lb]{\normalsize{\Black{$p_{4}$}}}
    \Line(101,30)(109,30)
    \Line(162,-21)(170,-21)
    \Line(109,40)(109,31)
    \Line(162,-21)(162,-30)
    \Line(103,37)(109,30)
    \Line(162,-21)(168,-28)
    \Line(162,30)(168,37)
    \Line(103,-28)(109,-21)
  \end{picture}
}
\end{center}
when $m_{1}^{2}=m_{3}^{2}=0$  and therefore
\begin{eqnarray}
\Delta_{2me} &=& s x_1 x_3+ u x_2 x_4 + m_{2}^2 x_{2} x_{3} + m_{4}^2 x_{4} x_{1} \\ \nonumber
\end{eqnarray}
Upon the following change of variables

\begin{eqnarray}
x_1 = \sigma_{1} (1 - \tau_{1}) \\ \nonumber
x_2 = \sigma_{1} \tau_{1} \\ \nonumber
x_3 = \sigma_{2} \tau_{2} \\ \nonumber
x_4 = \sigma_{2} (1 - \tau_{2}) \\ \nonumber
|\frac{\partial (x_{i})}{\partial (\sigma_{i}, \tau_{i})}|= \sigma_{1} \sigma_{2}
\end{eqnarray}
the integration over $\sigma_{i}$ factorizes and one gets

\begin{eqnarray}\label{ampans}
I^{2me}(p_{i},D_{IR},\mu_{IR})= (\mu_{IR}^2)^{\epsilon_{IR}} \Gamma(4 - \frac{D_{IR}}{2}) \int d \sigma_{1} d \sigma_{2} \sigma_{1}^{\frac{D_{IR}}{2}-3} \sigma_{2}^{\frac{D_{IR}}{2}-3} \delta( 1 - \sigma_{1} - \sigma_{2}) \\ \nonumber
\int_{0}^{1} d \tau_{1} d \tau_{2} \frac{1}{( - (s + u - m_{2}^2 -m_{4}^2)  \tau_{1} \tau_{2} + (u - m_{2}^2 ) \tau_{1} + (s - m_{2}^2) \tau_{2} + m_{2}^2)^{4 - \frac{D_{IR}}{2}}}
\end{eqnarray}
In this expression one can observe much similarity with the Wilson loop diagram.
Indeed, we will show further, that proper identification of parameters allows us
to connect it with Wilson loop diagram explicitly.

It is important that the special combinations of Feynman parameters play the role of parametrization of the point
in the Wilson polygon which emerges
in one-loop calculation
\begin{eqnarray} \label{wilsondef}
W({\cal C}_{n})=\frac{1}{N} Tr {\cal P} \exp [ i g \oint d \tau  \dot{x}^{\mu}(\tau) A_{\mu} (x(\tau))]
\end{eqnarray}

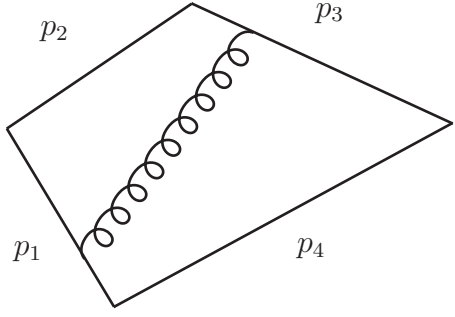
\begin{center}
\fcolorbox{white}{white}{
  \begin{picture}(168,129) (93,-44)
    \SetWidth{1.0}
    \SetColor{Black}
    \Line(134,-43)(94,24)
    \Line(94,24)(163,71)
    \Line(163,71)(260,26)
    \Line(260,26)(134,-43)
    \Gluon(123,-25)(186,60){4.5}{9}
    \Text(97,-26)[lb]{\normalsize{\Black{$p_{1}$}}}
    \Text(107,57)[lb]{\normalsize{\Black{$p_{2}$}}}
    \Text(210,64)[lb]{\normalsize{\Black{$p_{3}$}}}
    \Text(203,-25)[lb]{\normalsize{\Black{$p_{4}$}}}
  \end{picture}
}
\end{center}

We assume that $D_{UV}=d_{UV} - 2 \epsilon_{UV}$ , $p_{1}$ and $p_{3}$ are light-like,
$x_{1}= p_{1} \tau_{1}$ , $x_{2}= p_{1}+ p_{2} + p_{3} \tau_{2}$
and the standard propagator in the Feynman gauge
\begin{eqnarray} \label{propagator}
G^{F}_{\mu \nu}(x-y)= - \eta_{\mu \nu} \frac{(\pi \mu_{UV}^2)^{\epsilon_{UV}}}{4 \pi^2} \frac{\Gamma(\frac{D_{UV}}{2}-1)}{(-(x-y)^2 + i \epsilon)^{\frac{D_{UV}}{2}-1}}
\end{eqnarray}
Ignoring trivial factor $\frac{g^2 C_{F}}{16 \pi^2}$ we get the following expression for the diagram
\begin{eqnarray} \label{wilsonans}
& &I^{W}_{i j} (p_{i},D_{UV},\mu_{UV}) = \Gamma(\frac{D_{UV}}{2}-1) (\pi \mu_{UV}^2)^{\epsilon_{UV}} \\ \nonumber
& &\int_{0}^{1} d \tau_{i} d \tau_{j} \frac{m_{2}^2 + m_{4}^2 - s - u}{(- (s + u - m_{2}^2 -  m_{4}^2 ) \tau_{1} \tau_{2}  + \tau_{1} (s - m_{2}^2 ) + \tau_{2} (u - m_{2}^2) + m_{2}^2)^{\frac{D_{UV}}{2}-1}} \\ \nonumber
\end{eqnarray}
and from (\ref{ampans}) and (\ref{wilsonans}) we can make identification of the parameters to match two expressions. Namely substituting $\frac{D_{UV}}{2}-1 = 4 - \frac{D_{IR}}{2} $ we get

\begin{eqnarray}
d_{UV} + d_{IR} &=& 10 \\ \nonumber
\epsilon_{IR} &=& - \epsilon_{UV} \\ \nonumber
(\mu^2_{UV} \pi)^{\epsilon_{UV}} &=& (\mu^2_{IR})^{\epsilon_{IR}}
\end{eqnarray}
and the exact correspondence  reads as follows
\begin{eqnarray}\nonumber
I^{2me}(p_{i},D_{IR},\mu_{IR}) &=& \frac{1}{m_{2}^2 + m_{4}^2 - s - u} \int_{0}^{1} d \sigma \sigma^{2 - \frac{D_{UV}}{2}} (1 - \sigma)^{2 - \frac{ D_{UV}}{2}} I^{W}_{i j} (p_{i},D_{UV},\mu_{UV}) \\ \nonumber
&=& \frac{1}{m_{2}^2 + m_{4}^2 - s - u} \frac{\Gamma (3 - \frac{D_{UV}}{2})^2}{\Gamma (6 - D_{UV})}  I^{W}_{i j} (p_{i},D_{UV},\mu_{UV})
\end{eqnarray}

Note that it is possible to represent the expression for the Wilson polygon
in the form which involves a kind of the integral over the reparametrization of the boundary contour.
\begin{eqnarray}\nonumber
I^{2me}(p_{i},D_{IR},\mu_{IR}) &=& \frac{1}{m_{2}^2 + m_{4}^2 - s - u} \int_{0}^{1} d \sigma I^{W} ( {\cal C}(\sigma),p_{i},D_{UV},\mu_{UV})\\ \nonumber
&&{\cal C}(\sigma): p_{i} \rightarrow \sqrt{\sigma (1 - \sigma)} p_{i} \\ \nonumber
\end{eqnarray}
Such form of the answer was suggested at the strong coupling \cite{polyakov}
when the integration over the reparametrizations of the boundary of the Wilson loop
is necessary to restore the conformal invariance of the answer.

Suppose we are interested in $D_{UV}=4 - 2 \epsilon$ Wilson loop diagram. Than, using the known connection between Wilson diagram
and finite part of the box, we get
\begin{eqnarray}
I^{W}_{i j} (p_{i},4 - 2 \epsilon) &=& (m_{2}^2 + m_{4}^2 - s - u) \frac{ \Gamma (2 + 2 \epsilon)}{\Gamma^2 (1 + \epsilon)} I^{2me}(p_{i},6 + 2 \varepsilon) \\ \nonumber
&=& Fin [\frac{ \Gamma (1 + 2 \epsilon)}{\Gamma^2 (1 + \epsilon)} I^{2me}(p_{i}, 4 + 2 \epsilon) \frac{1}{2} (m^2_{2} m^2_{4} - s u)]
\end{eqnarray}
and therefore the following relation  provides the desired duality

\begin{eqnarray}
  I^{2me}(p_{i},6 + 2 \epsilon) &=& Fin [\frac{I^{2me}(p_{i}, 4 + 2 \epsilon)}{1 + 2 \epsilon}  \frac{(s u - m^2_{2} m^2_{4})}{2 (s + u - m_{2}^2 -  m_{4}^2)}]
\end{eqnarray}

Such connection between $D$ and $D-2$ dimensional scalar loop integrals indeed exists and goes to the papers \cite{DavydRed}.
Here we are interested in the case of $D=6$ two-mass easy boxes and their connection with $D=4$ ones \cite{Nizic2}.
The formula reads as follows (see appendix A)
\begin{eqnarray}\nonumber
I^{2me}(6 + 2 \epsilon) = \frac{1}{(1 + 2 \epsilon) z_{0}} (I^{2me}(4 + 2 \epsilon) - \sum_{i=1}^{4} z_{i} I^{2me}(4 + 2 \epsilon; 1 - \delta_{k i} ) )
\end{eqnarray}
where
\begin{eqnarray}\nonumber
z_{0}= \sum_{i=1}^{4} z_{i} &=& 2 \frac{s + u - m_{2}^2 - m_{4}^{2}}{s u - m^2_{2} m^2_{4}} \\ \nonumber
z_{1} &=& \frac{u-m_{2}^{2}}{s u - m^2_{2} m^2_{4}} \\ \nonumber
z_{2} &=& \frac{s-m_{4}^{2}}{s u - m^2_{2} m^2_{4}} \\ \nonumber
z_{3} &=& \frac{u-m_{4}^{2}}{s u - m^2_{2} m^2_{4}} \\ \nonumber
z_{4} &=& \frac{s-m_{2}^{2}}{s u - m^2_{2} m^2_{4}} \\ \nonumber
\end{eqnarray}
As can be easily seen the $\sum_{i=1}^{4} z_{i} I^{4}(4 + 2 \epsilon; 1 - \delta_{k i})$ does precisely the job of taking  the finite part.


As we know from the calculation of the one-loop NMHV amplitudes \cite{Sok1}
the new ingredients emerge namely two-mass hard and
three-mass boxes. Thus if one wants to extend the duality between Wilson loop
and amplitudes to NMHV case one should be able to get these ingredients from
Wilson loop language.

In the case of two-mass easy box the structure of function in the space of Feynman parameters
space allowed us using change of variables to get the Wilson loop diagram multiplied
by the simple numerical integral. We can interpret this integral as an integral over the reparameterizations
of the contours.
One can try to use the same approach of splitting Feynman parameters in two pairs:
one pair parameterizes the contour while the second  yields the standard parametrization
of points where gluon propagator is attached.

For more complicated cases  than two-mass easy box
the factorization fails and therefore the simple geometrical interpretation
does not work.
Namely if we make all  legs massive in the Feynman box and consider the corresponding
Wilson contour the integrands in the amplitude and the Wilson loop looks as  follows:

\begin{eqnarray}
&\Delta_{W} = - (s + u - m_{2}^2 -m_{4}^2)  \tau_{1} \tau_{2} + (u - m_{2}^2 ) \tau_{1} + (s - m_{2}^2) \tau_{2} + m_{2}^2 &\\ \nonumber
&- m_{1}^2 \tau_{1} (1 - \tau_{1}) - m_{3}^2 \tau_{2} (1 - \tau_{2}) &\\ \nonumber
\end{eqnarray}

\begin{eqnarray}
&\Delta_{A} = \sigma_{1} \sigma_{2} [ - (s + u - m_{2}^2 -m_{4}^2)  \tau_{1} \tau_{2} + (u - m_{2}^2 ) \tau_{1} + (s - m_{2}^2) \tau_{2} + m_{2}^2 ] & \\ \nonumber
&+ m_{1}^2 \sigma_{1}^2 \tau_{1} (1 - \tau_{1}) + m_{3}^2 \sigma_{2}^2 \tau_{2} (1 - \tau_{2}) & \\ \nonumber
&=  \sigma_{1} \sigma_{2} \Delta_{W}+ m_{1}^2 \sigma_{1} \tau_{1} (1 - \tau_{1}) + m_{3}^2 \sigma_{2} \tau_{2} (1 - \tau_{2}) &\\ \nonumber
\end{eqnarray}
We have not found simple geometrical interpretation of transformation from $\Delta_{W}$ to $\Delta_{A}$
in terms of the reparametrizations of the Wilson contour
and we can not naturally connect two-mass hard and harder boxes diagrams with Wilson diagrams
for correspondent contours.
That is if the connection between NMHV amplitudes and Wilson polygon-like objects exists which is
expected according to the
T-dual picture of $AdS_{5} \times S_{5}$ superstring \cite{ber} than it seems to be  more
complicated.

\section{3-point function - Wilson triangle duality }

In this section we consider the example of the similar duality
for the 3-point function and it will be clear how the
generalization of the duality for the "two-mass hard" diagram
involves the particular vertex operator. To start with let
us mention also interesting relation between the one-loop
3-point amplitude and the two-loop vacuum energy in the scalar
theory. Namely if one considers the 3-point function $I(p_1^2, p_2 ^2, p_3 ^2)$
with the
the external virtualities $p_1^2, p_2 ^2, p_3 ^2$ and the two-loop
vacuum energy $J(m_1 ^2, m_2 ^2, m_3 ^2)$ with the masses of the
three internal propagators
$m_1 ^2, m_2 ^2, m_3 ^2$ then the following relation holds \cite{tausk}
\beq
I(D=4-2\epsilon, p_1^2, p_2 ^2, p_3 ^2)=J(4 +2\epsilon, m_1 ^2, m_2 ^2, m_3 ^2)
\eeq
That is the duality discussed below can be applied both for the
one-loop amplitude and the two-loop vacuum energy.

Consider the most general triangle in the massless scalar theory

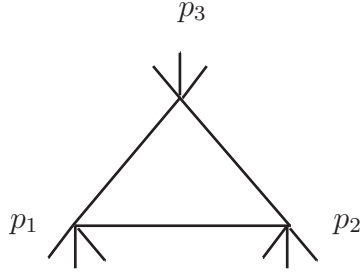
\begin{center}
\fcolorbox{white}{white}{
  \begin{picture}(153,115) (92,-15)
    \SetWidth{1.0}
    \SetColor{Black}
    \Line(112,2)(152,50)
    \Line(152,50)(193,2)
    \Line(193,2)(113,2)
    \Line(152,67)(152,51)
    \Text(89,-1)[lb]{\normalsize{\Black{$p_{1}$}}}
    \Text(210,-1)[lb]{\normalsize{\Black{$p_{2}$}}}
    \Text(152,79)[lb]{\normalsize{\Black{$p_{3}$}}}
    \Line(113,2)(113,-14)
    \Line(192,2)(192,-14)
    \Line(152,50)(142,62)
    \Line(203,-11)(193,1)
    \Line(124,-11)(114,1)
    \Line(153,50)(162,62)
    \Line(103,-10)(112,2)
    \Line(183,-10)(192,2)
  \end{picture}
}
\end{center}

In the Feynman parametrization it is equal to

\begin{eqnarray}
p_{1}+p_{2}+p_{3} &=& 0 \\ \nonumber
I_{\triangle} (p_{i},D_{IR},\mu_{IR}) &=&- (\mu_{IR}^2)^{\epsilon_{IR}} \Gamma(3 - \frac{D_{IR}}{2}) \int \prod d x_{i} \frac{\delta( 1 - x_1 - x_2 - x_3)}{(- \Delta)^{3 - \frac{D_{IR}}{2}}} \\ \nonumber
\Delta &=& m_{3}^2 x_{1} x_{2} + m_{2}^2 x_{1} x_{3} + m_{1}^2 x_{2} x_{3} \\ \nonumber
\end{eqnarray}
and assuming $p_{3}^2=0$  we have

$$\Delta = m_{2}^2 x_{1} x_{3} + m_{1}^2 x_{2} x_{3} $$
Let us make the following change of variables

\begin{eqnarray}
x_1 = \sigma (1 - \tau) \\ \nonumber
x_2 = \sigma \tau \\ \nonumber
\end{eqnarray}

which amounts to
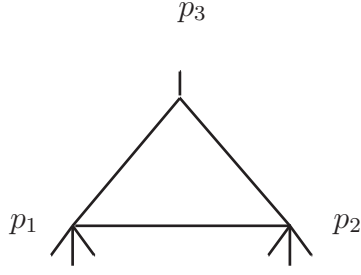
\begin{center}
\fcolorbox{white}{white}{
  \begin{picture}(153,114) (92,-15)
    \SetWidth{1.0}
    \SetColor{Black}
    \Line(112,1)(152,49)
    \Line(152,49)(193,1)
    \Line(193,1)(113,1)
    \Line(152,59)(152,50)
    \Text(89,-2)[lb]{\normalsize{\Black{$p_{1}$}}}
    \Text(210,-2)[lb]{\normalsize{\Black{$p_{2}$}}}
    \Text(152,78)[lb]{\normalsize{\Black{$p_{3}$}}}
    \Line(112,1)(112,-14)
    \Line(120,-10)(112,1)
    \Line(104,-10)(112,1)
    \Line(193,1)(193,-14)
    \Line(201,-10)(193,1)
    \Line(185,-10)(193,1)
  \end{picture}
}
\end{center}

\begin{eqnarray}
I_{\triangle} (p_{i},D_{IR},\mu_{IR}) &=& (\mu_{IR}^2)^{\epsilon_{IR}} \Gamma(3 - \frac{D_{IR}}{2}) \int d \sigma d x_{3} \sigma  \frac{\delta( 1 - \sigma - x_3)}{(\sigma x_{3})^{3 - \frac{D_{IR}}{2}}}\\ \nonumber
& & \int_{0}^{1} d \tau \frac{1}{(m_{2}^2 (1 - \tau) + m_{1}^2 \tau )^{3 - \frac{D_{IR}}{2}}} \\ \nonumber
\end{eqnarray}
In Wilson-dual language we can interpret it in the following way

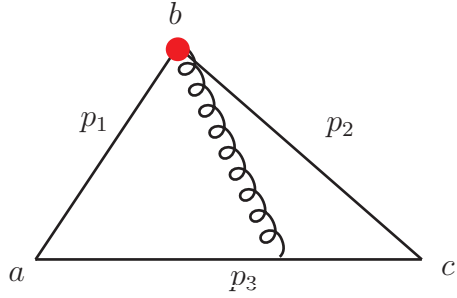
\begin{center}
\fcolorbox{white}{white}{
  \begin{picture}(194,117) (58,-22)
    \SetWidth{1.0}
    \SetColor{Black}
    \Line(65,-15)(118,65)
    \Line(118,65)(209,-15)
    \Line(65,-15)(209,-15)
    \Gluon(118,65)(156,-14){3.5}{9}
    \SetColor{Red}
    \Vertex(118,64){4.472}
    \Text(55,-23)[lb]{\normalsize{\Black{$a$}}}
    \Text(115,74)[lb]{\normalsize{\Black{$b$}}}
    \Text(217,-21)[lb]{\normalsize{\Black{$c$}}}
    \Text(82,33)[lb]{\normalsize{\Black{$p_{1}$}}}
    \Text(174,31)[lb]{\normalsize{\Black{$p_{2}$}}}
    \Text(138,-27)[lb]{\normalsize{\Black{$p_{3}$}}}
  \end{picture}
}
\end{center}

\begin{eqnarray}
m_{2}^2 (1 - \tau) + m_{1}^2 \tau & = & (p_{2} + p_{3} \tau)^{2}\\ \nonumber
\end{eqnarray}
and the  identification of parameters reads as follows

\begin{eqnarray}
d_{UV} + d_{IR} &=& 8 \\ \nonumber
\epsilon_{IR} &=& - \epsilon_{UV} \\ \nonumber
(\mu^2_{UV} \pi)^{\epsilon_{UV}} &=& (\mu^2_{IR})^{\epsilon_{IR}}
\end{eqnarray}

Therefore this diagram can be understood assuming the presence of the
vertex operator
\begin{eqnarray}
< Tr {\cal P} q^{\mu} A_{\mu} (x_{b}) \exp [ i g \oint_{{\cal C}} d \tau  \dot{x}^{\mu}(\tau) A_{\mu} (x(\tau))] >
\end{eqnarray}
where $q^{\mu}$ can be chosen as be arbitrary vector which is not orthogonal to $p_{3}$ in
Minkowski sense,  $(p_{3} q) \neq 0$.
This $q^{\mu}$ can be naturally identified with the polarization vector of
correspondent external gluon. Hence we have an example of possible
extension of Wilson dual side, when it becomes sensitive to polarizations
of external gluons. This example provides some intuition for the
possible generalization of the duality to less symmetric theories or
NMHV amplitudes.
Nevertheless, the problem of the interpretation the three-mass triangle and all boxes
harder than two-mass easy one in terms of Wilson loop diagrams is still open.

\section{Analytical structure of light-like Wilson loop}
\subsection{General comments}
In this section we discuss the analytical structure
of light-like Wilson loop. If the correspondence between
MHV amplitudes and Wilson loops is true at any order of
perturbation theory  obviously their analytical
structure namely the location of singularities, branches and discontinuities
in the space of kinematic moduli should match each other.
Thus there emerges two interesting problems on its own: analytical
structure of perturbative light-like Wilson loop and the similar
question concerning the areas in $AdS_{5}$ bounded by the light-like contour.

Here we start analysis of analytical structure of perturbative
Wilson loop.
Firstly, we can do it using its connection with scattering
amplitudes. The fact of unitarity of QFT
leads to optical theorem and allows one to take different branch
cuts and develop generalized unitarity method to simplify
loop computations. Using the correspondence between Wilson loops
and amplitudes we can reformulate optical theorem at one loop
 in terms of Wilson loops.

Secondly, one can analyze the analytical structure of
every Wilson diagram on its own. The systematic method of clarifying the
structure of
singularities of Feynman amplitudes was developed long time ago
in the theory of analytic S-matrix. It can be obviously
applied to Wilson loop diagrams.
At one-loop level using the results of previous section we can apply
Cutkosky rules to $10 - D_{W}$ boxes which are dual to Wilson diagrams
to get the result for given diagram while at higher
orders the additional arguments are required.

\subsection{Landau singularities for the Wilson loop}

Consideration here is parallel to \cite{Lowe}, where excellent introduction
to the problem can be found.
Suppose we deal with scalar massless theory Feynman integrals in $D$ dimensions than we have
for any diagram \cite{Smirnov}:

\begin{eqnarray} \label{repr3}
 I \simeq \int_{0}^{1} \prod d \alpha_{i} \delta (1 - \sum_{i} \alpha_{i}) \frac{{\cal U}^{N - (L+1) \frac{D}{2}} }{(- {\cal V})^{N - L \frac{D}{2}}}
\end{eqnarray}
Here: $L$ - number of loops; $N$ - number of propagators; $\alpha_{i}$ corresponds to the $i$-th propagator in the diagram of the form $\frac{1}{q_{i}^2}$.

${\cal U} = \sum_{T \in T_{1}} \prod_{i \in \bar{T}} \alpha_{i}$ - sum over so-called 1-trees, degree $L$ in $\alpha$.\\

${\cal V} = \sum_{T \in T_{2}} \prod_{i \in \bar{T}} \alpha_{i} (Q_{T})^2$- sum over so-called 2-trees, degree $(L+1)$ in $\alpha$.\\
The $I$ can be considered as the function of complex kinematical parameters
and the natural question arises: where its
singularities in the space of parameters are located?
The answer to this question is given by the Landau equations which can be written
in the following form:
       $$
           \left\{
           \begin{array}{rcl}
            \frac{\partial {\cal V}}{ \partial \alpha_{i}}  =  0  & \wedge & \alpha_{i} = 0 \\
              {\cal V}  & = & 0 \\
           \end{array}
           \right.
       $$

The same analysis can be applied for any particular Wilson loop. If one
considers the family of more simple diagrams where every propagator has one leg
lying on the boundary, than consideration is in full analogy with the case of amplitudes.
Namely the diagram has the following structure
\begin{eqnarray} \label{wrepr1}
 W \simeq \int_{0}^{1} \prod_{i=1}^{2 N} d \tau_{i} \Theta_{Path} ( x(\tau_{\sigma_{1}}) > x(\tau_{\sigma_{2}}) > ... > x(\tau_{\sigma_{2 N}})) \prod_{k=1}^{V_{3}} \hat{L}_{k} \frac{d^{D} z_{1} d^{D} z_{2} ... d^{D} z_{V}}{\prod_{k = 1}^{N} ( - x_{k}^{2} )^{\frac{D}{2}-1} }
\end{eqnarray}
Here $\hat{L}_{k}$ is the differential operator independent of $z_{i}$'s, which comes from three-gluon vertexes \cite{Brand2}; $V_{3}$ - number
of three-gluon vertexes
\begin{eqnarray} \label{vertexop}
&A^{\mu_{1}} A^{\mu_{2}} A^{\mu_{3}} \int d^{D} z_{k} Tr[ \partial_{\mu} ( A_{\nu} [A^{\mu}, A^{\nu}] ) (z_{k})& \\ \nonumber
&\sim [ \eta^{\mu_{1} \mu_{2}} (\partial_{1}^{\mu_{3}} - \partial_{2}^{\mu_{3}}) + \eta^{\mu_{2} \mu_{3}} (\partial_{1}^{\mu_{1}} - \partial_{2}^{\mu_{1}}) + \eta^{\mu_{1} \mu_{3}} (\partial_{1}^{\mu_{2}} - \partial_{2}^{\mu_{2}})] G(x_{1},x_{2},x_{3})& \\ \nonumber
& \hat{L}_{k}^{\mu_{1} \mu_{2} \mu_{3}} G(x_{1},x_{2},x_{3})&\\ \nonumber
\end{eqnarray}
Than if the points $(x_{1},x_{2},x_{3})$ lie on the edges $(y_{1},y_{2},y_{3})$
\begin{eqnarray} \label{vertex}
&\hat{L}_{k} =\dot{y}_{1\mu_{1}} \dot{y}_{2\mu_{2}} \dot{y}_{3\mu_{3}}  \hat{L}_{k}^{\mu_{1} \mu_{2} \mu_{3}} & \\ \nonumber
\end{eqnarray}
For any ordering, there exists the change of variables of integration with the Jacobian $J$ independent on the kinematical variables that makes the simple integration interval:
\begin{eqnarray} \label{repr3}
\int_{0}^{1} \prod_{i=1}^{2 N} d \tau_{i} \Theta_{Path} (\tau) \rightarrow \int_{0}^{1} \prod_{i=1}^{2 N} d \tilde{\tau}_{i} J(\tilde{\tau})
\end{eqnarray}
 If we have the following ordering along one of the edges $\int_{0}^{1} d \tau_{n} \int_{0}^{\tau_{n}} d \tau_{n-1} ...\int_{0}^{\tau_{2}} d \tau_{1}$. Than one can choose
\begin{eqnarray}
\tau_{n} &=& \tilde{\tau}_{n} \\ \nonumber
\tau_{n-1} &=& \tau_{n}  \tilde{\tau}_{n-1} = \tilde{\tau}_{n} \tilde{\tau}_{n-1} \\ \nonumber
&...& \\ \nonumber
\tau_{1} &=& \tau_{2}  \tilde{\tau}_{1} = \tilde{\tau}_{n} \tilde{\tau}_{n-1} ... \tilde{\tau}_{1} \\ \nonumber
J(\tilde{\tau}) &=& \prod_{j=1}^{n} \tilde{\tau}_{j}^{j - 1} \\ \nonumber
\end{eqnarray}
If the vertexes are absent than the Landau equations take the form
       $$
           \left\{
           \begin{array}{rcl}
            \frac{\partial (\sum \alpha_{k} x_{k}^{2}) }{ \partial \alpha_{i} }  =  0  & \wedge & \alpha_{i} = 0 \\
            \frac{\partial (\sum \alpha_{k} x_{k}^{2}) }{ \partial \tilde{\tau}_{i} }  =  0  & \wedge & \tilde{\tau}_{i} = 0 \wedge \tilde{\tau}_{i} = 1 \\
            \sum \alpha_{k} x_{k}^{2}  & = & 0 \\
           \end{array}
           \right.
       $$
In the presence of vertexes we can introduce Feynman parameters
\begin{eqnarray} \label{repr2}
 W \simeq \int_{0}^{1} \prod_{i=1}^{2 N} d \tilde{\tau}_{i} J(\tilde{\tau}) \prod_{k=1}^{V_{3}} \hat{L}_{k}  \int_{0}^{1} \prod_{k=1}^{N} d \alpha_{k} \alpha_{k}^{\frac{D}{2}-2} \delta (1 - \sum_{i} \alpha_{i}) \frac{d^{D} z_{1} d^{D} z_{2} ... d^{D} z_{V}}{ [- \sum \alpha_{k} x_{k}^{2}]^{N (\frac{D}{2}-1)}}
\end{eqnarray}
and integration over the vertex position could be done yielding the answer
\begin{eqnarray} \label{repr3}
 W \simeq \int_{0}^{1} \prod_{i=1}^{2 N} d \tilde{\tau}_{i} J(\tilde{\tau}) \prod_{k=1}^{V_{3}} \hat{L}_{k} \int_{0}^{1} \prod d \alpha_{k} \alpha_{k}^{\frac{D}{2}-2} \delta (1 - \sum_{i} \alpha_{i}) \frac{{\cal U}_{W}(\alpha_{i})}{(- {\cal V}_{W})^{(N - V)\frac{D}{2} - N}}
\end{eqnarray}
Here: $V$ - number of vertexes; $N$ - number of propagators; $\alpha_{i}$ corresponds to the $i$-th propagator in the diagram of the form $\frac{1}{(-x_{i}^2)^{\frac{D}{2}-1}}$; ${\cal U}_{W}$ and ${\cal V}_{W}$ - the result of the integration over the loop momenta.



Finally we get the following Landau equations

       $$
           \left\{
           \begin{array}{rcl}
            \frac{\partial {\cal V}_{W}}{ \partial \alpha_{i} }  =  0  & \wedge & \alpha_{i} = 0 \\
            \frac{\partial {\cal V}_{W}}{ \partial \tilde{\tau}_{i} }  =  0  & \wedge & \tilde{\tau}_{i} = 0 \wedge \tilde{\tau}_{i} = 1 \\
            {\cal V}_{W}  & = & 0 \\
           \end{array}
           \right.
       $$

\subsection{Imaginary part of the Wilson loop at one loop}

In unitary theory one can exploit the unitarity of $S$-matrix to
get the following identity

$$S^{+}S=1$$
$$S=1 + i T$$

\begin{eqnarray} \label{sam}
2 Im ({\cal A}(in \rightarrow out)) = - i ({\cal A}(in \rightarrow out)+{\cal A}^{*} (out \rightarrow in)) \\ \nonumber
= \sum_{states} {\cal A}^{*}(out \rightarrow all) {\cal A}(in \rightarrow all)
\end{eqnarray}

We are interested in amplitudes with $n$ outgoing particles. In that case the RHS sum becomes the
sum over state with integration  over a Lorentz invariant phase space corresponding to the final
particles. Pushing this statement to diagrammatic level one ends with Cutkosky rules and prescription
of cutting propagators:

\begin{eqnarray} \label{sam}
\frac{1}{k^2 + i \epsilon} \rightarrow \Theta (k_{0}) \delta(k^2)
\end{eqnarray}
It is well-known that to get the imaginary part of the diagram to given order one should sum over all possible
cuts of all diagrams and over all possible intermediate states. Than one should do the integration over LIPS.
The Wilson loop knows about ${\cal N}=4$ SYM
particle content only through correction to the gluon propagator and
the vertices. At one loop level it is obviously insensitive to
particle content.
On the amplitudes side the cut is on the contrary essentially dependent
on the particle content and tree-level amplitude even at one loop. According to the strong version of
the correspondence which is true at one loop

\begin{eqnarray} \label{corr}
  {\cal A}_{n}^{MHV}= {\cal A}_{n}^{tree} W({\cal C}_{n})
\end{eqnarray}
with necessary identification of parameters. Since ${\cal A}_{n}^{tree}$ is rational function of kinematical
variables it  does not contribute to the cut. Hence we can rewrite optical theorem
as
\begin{eqnarray} \label{sam}
2 Im [ W({\cal C}_{n}) ]= 2 Im (\frac{{\cal A}(in \rightarrow out)}{{\cal A}^{tree}(in \rightarrow out)}) \\ \nonumber
= \sum_{states} \frac{{\cal A}^{*}(out \rightarrow all) {\cal A}(in \rightarrow all)}{{\cal A}^{tree}(in \rightarrow out)} \\ \nonumber
\end{eqnarray}

At one loop level we  have sum over products of tree-level amplitudes. Denoting this sum  divided by ${\cal A}_{n}^{tree}$
as $V_{W}$ (which can be found in appendix B) we have
\begin{eqnarray} \label{optw}
 Im [ W({\cal C}) ]= \int_{{\cal C}_{L}{\cal C}_{R}}  V_{W} ({\cal C}_{L},{\cal C}_{R})
\end{eqnarray}

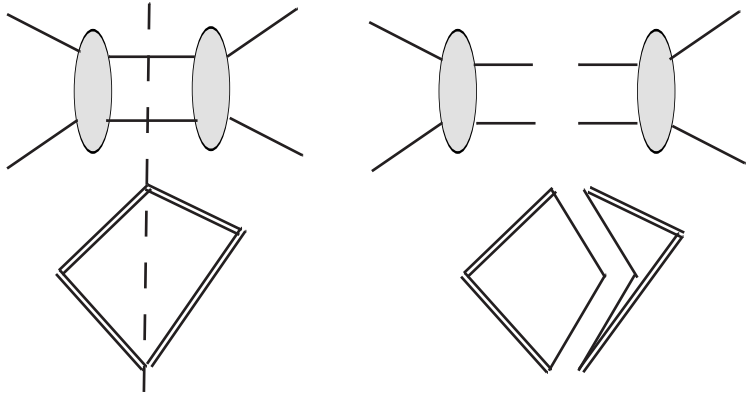
\begin{center}
\fcolorbox{white}{white}{
  \begin{picture}(277,148) (91,-20)
    \SetWidth{1.0}
    \SetColor{Black}
    \GOval(124,94)(23,7)(0){0.882}
    \GOval(168,95)(23,7)(0){0.882}
    \Line(130,107)(162,107)
    \Line(129,83)(163,83)
    \Line[dash,dashsize=10](145,127)(143,-19)
    \Line[double,sep=2](143,-10)(111,26)
    \Line[double,sep=2](112,26)(145,58)
    \Line[double,sep=2](144,58)(179,41)
    \Line[double,sep=2](180,42)(145,-9)
    \Line(119,109)(92,123)
    \Line(202,70)(175,84)
    \Line(118,83)(92,65)
    \Line(200,125)(174,107)
    \GOval(260,94)(23,7)(0){0.882}
    \GOval(334,94)(23,7)(0){0.882}
    \Line(254,106)(227,120)
    \Line(254,82)(228,64)
    \Line(367,67)(340,81)
    \Line(365,124)(339,106)
    \Line(266,104)(288,104)
    \Line(267,82)(289,82)
    \Line(305,104)(327,104)
    \Line(305,82)(327,82)
    \Line[double,sep=2](294,-11)(262,25)
    \Line[double,sep=2](263,25)(296,57)
    \Line(295,57)(315,24)
    \Line(315,25)(294,-11)
    \Line[double,sep=2](309,57)(344,40)
    \Line[double,sep=2](343,40)(306,-11)
    \Line(307,57)(327,24)
    \Line(326,25)(305,-11)
  \end{picture}
}
\end{center}
where integration goes over contours which one could get by breaking the loop into two parts, inserting special vertex, which
one could find from summing over states in ${\cal N}=4$ SYM and then by integration over contours which are limited by momentum conservation and light-like condition for every edge.

On the other hand the problem of finding imaginary part can be considered at diagrammatical level
where the connection with the box in dual dimension makes it possible to apply Cutkosky rules.
Of course, on this way there is no any summation over states. It would be nice to understand
how the vertices from the dual amplitude picture occur in the game.
The following picture arises if one considers the quadruple cut

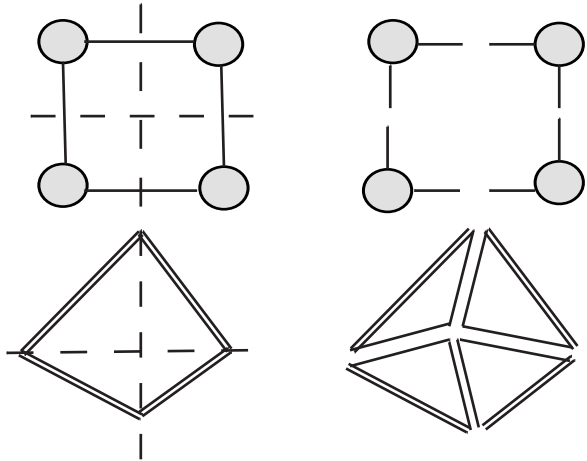
\begin{center}
\fcolorbox{white}{white}{
  \begin{picture}(217,173) (152,-143)
    \SetWidth{1.0}
    \SetColor{Black}
    \Line[double,sep=2](282,-101)(326,-56)
    \Line(327,-56)(318,-92)
    \Line(318,-91)(281,-101)
    \Line[double,sep=2](325,-131)(280,-107)
    \Line(318,-97)(281,-107)
    \Line(318,-98)(326,-131)
    \Line[double,sep=2](332,-56)(365,-100)
    \Line(332,-56)(323,-92)
    \Line(323,-92)(364,-100)
    \Line[double,sep=2](364,-104)(331,-130)
    \Line(321,-97)(329,-130)
    \Line(322,-97)(363,-105)
    \GOval(295,-41)(8,9)(0){0.882}
    \Line(295,-17)(295,-33)
    \Line(305,-41)(322,-41)
    \GOval(297,15)(8,9)(0){0.882}
    \GOval(359,14)(8,9)(0){0.882}
    \GOval(359,-38)(8,9)(0){0.882}
    \Line(296,6)(296,-10)
    \Line(359,6)(359,-10)
    \Line(359,-14)(359,-30)
    \Line(333,-41)(350,-41)
    \Line(333,14)(350,14)
    \Line(306,14)(323,14)
    \Line[dash,dashsize=10](203,-56)(203,-142)
    \Line[dash,dashsize=10](153,-102)(243,-101)
    \Line[double,sep=2](203,-126)(158,-102)
    \Line[double,sep=2](159,-102)(203,-57)
    \Line[double,sep=2](203,-57)(236,-101)
    \Line[double,sep=2](236,-101)(203,-125)
    \GOval(174,15)(8,9)(0){0.882}
    \GOval(174,-39)(8,9)(0){0.882}
    \GOval(234,-40)(8,9)(0){0.882}
    \GOval(232,14)(8,9)(0){0.882}
    \Line(174,7)(175,-31)
    \Line(233,6)(234,-32)
    \Line(182,15)(223,15)
    \Line(182,-41)(224,-41)
    \Line[dash,dashsize=10](203,29)(203,-47)
    \Line[dash,dashsize=10](162,-13)(245,-13)
  \end{picture}
}
\end{center}
It is interesting to note the role of  the coefficient
\beq
{\cal C}^{2me}_{MHV}= \delta^{8} (\sum_{i=1}^{n} \lambda_{i} \eta_{i}) \Delta
\eeq
which appears from quadruple cut of the MHV amplitude and is defined by the structure of tree amplitudes in ${\cal N}=4$ SYM.
In the Wilson loop calculation which is blind to trees it appears while we go down from $D=6$ to $D=4$ dimensions,
namely
\beq
z_{0} \sim \frac{1}{\Delta}
\eeq

\section{On the geometry of UV/IR divergences}

Let us discuss the  interpretation of the
divergent contributions. The IR  singularity
of the amplitude corresponds  to the UV singularity of
the  cusps hence the very issue of the
proper IR regularization of the amplitude is essentially
related to the smoothing of the cusps in the polygon in the momentum space.

Let us make a few comments concerning the proper identification
of the cusp anomaly in the geometrical terms \cite{bgk,gorsky}. Since the amplitude
is expressed in terms of the hyperbolic volumes and area in 3D $AdS$ space
it is natural to question what the cusp anomaly corresponds to in the
same setting. That is we can start with the box with all external momenta
off-shell which is finite. Then approaching on-shell limit for two external momenta
the geometrical volume and area start to diverge which corresponds to the
divergence of the Feynman diagram.  Nevertheless we expect that the initial
geometry is partially seen in the divergent terms.

Recall that
$\Gamma_{cusp}(\theta,\alpha)$ is the cusp anomalous dimension
which for the cusp angle $\theta$ at one loop behaves as

\beq
\Gamma_{cusp}(\theta,\alpha)=\frac{\alpha C_{F}}{\pi}(\theta
\coth \theta -1)+ O(\alpha^2)
\eeq
It turns out \cite{bgk} that one-loop expression is nothing but the transition
amplitude in $AdS_3$
\beq
\Gamma_{cusp}(\theta)\propto <v'|1/\Delta_{S_3}|v>
\eeq
where two light-like vectors v and v' cross at the angle, and
$\Delta_{S_3}$ is the corresponding Laplace operator on the $SU(2)$ group manifold.
That is the one-loop anomaly
can be attributed to the amplitude along the  single edge of the basic simplex
upon the analytic continuation \cite{bgk}.
Note that these geodesics connecting two vertices are dressed by the quadratic
fluctuations.

Since the quantum geometry of the $AdS_3$ is governed by the $SL(2,C)$ Chern-Simons
theory the corresponding Wilson loop is just the particle moving in this background.
It is also possible to make the link with the $AdS_2$ geometry since the
one-loop cusp anomaly can be written as the wave functional in the
two dimensional YM theory on the disc integrated over its area.
\beq
\Gamma_{cusp}(\theta)\propto \int dA (Z(U,A) -Z(U,0))
\eeq
where $A$ is the area of the disc, $U$ is the boundary holonomy and  $Z(U,A)$
is partition function of the $2D$ YM theory in the disc geometry.

Since it is expected that the reparametrization of the boundary
enters the answer
it is natural to search for the Liouville interpretation of the cusp
anomaly. Contrary to the finite contribution where the reparametrization
part decouples and does not depend on the kinematical invariants we expect
that divergent "Liouville" contribution has nontrivial kinematical dependence.

The possible arguments which however deserve more justification look as follows \cite{gorsky}.
Consider two dimensional scalar field theory  with the equation of
motion

\beq (\partial_{t}^2 -\partial_{x}^2)\phi +m^2\phi=0 \eeq
whose solution has the following mode expansion

\beq \phi(x,t)=\int \frac{d\beta}{2\pi}
(a^{*}(\beta)e^{-i m (x \sinh \beta -t \cosh \beta} +
a(\beta)e^{i m (x \sinh\beta -t \cosh\beta}) \eeq
It is convenient to introduce Rindler coordinates

\ba x=r \cosh \theta,\qquad t=r \sinh \theta
\nonumber\\
-\infty < \theta < +\infty \qquad 0 < r <+\infty \ea in the
space-time region $x>|t|>0$. Upon the following Laplace transform
with respect to the radial coordinate

\beq \lambda_{\theta}(\alpha)=\int dr
e^{i m r \sinh \alpha}(-\frac{1}{r}\partial_{\theta} +
i m \cosh \alpha)\phi(r,\theta) \eeq
the commutation relation for  the Laplace transformed field reads as
\beq [\lambda(\alpha_1), \lambda(\alpha_2)]=i \hbar \tanh(\alpha_1
-\alpha_2)/2 \eeq
and the Hilbert space is spanned by vectors $a(\beta_n)\dots
a(\beta_1)|vac>$ where vacuum state is defined as \beq
a(\beta)|vac>=0 \qquad <vac|a^{+}(\beta)=0 \eeq

One can introduce two-point function  on the
"rapidity plane"
\beq F(\alpha_1-\alpha_2)=
<vac|\lambda(\alpha_1)  \lambda(\alpha_2)|vac>
\eeq
and explicit calculation amounts to the following answer \cite{luk}

\beq F(\alpha -i\pi)= - \frac{1}{\pi}\alpha /2 \coth(\alpha /2) +
singular \quad terms
\eeq
Hence the singular terms cancel in the
difference $ F(\alpha -i\pi) - F( -i\pi)$ which coincides with the
cusp anomaly in agreement with the interpretation of \cite{bgk} in
the first quantized picture.

The relation with the Liouville model becomes clear upon
the proper limiting procedure. To this aim we can try to
represent Klein- Gordon equation of motion as the zero curvature
condition for $SL(2,R)$ connection. Similarly  the equation
of motion in the
Liouville model
\beq (\partial_t^2 -\partial_x^2)\phi +
\frac{m^2}{b}e^{b\phi}=0
\eeq
can be considered as zero curvature condition for
$SL(2,R)$ valued connection $A_{\theta},A_{r}$. It is
convenient to
introduce the  monodromy matrix in the Liouville model
\beq
{\bf{T}}^{\theta}(\alpha) \propto e^{ i m R \sinh \alpha
\sigma_3} {\cal P} \exp(\int dr A_{r}(r,\theta,\alpha) ) \eeq
where R is cutoff,
which defines $\lambda_{Liov}(x)$ via relation
\beq \lambda(\alpha)
=-i \ln T_{11}(\alpha) \eeq
The latter reduces to the corresponding Klein-Gordon
function involved into the cusp anomaly and in
the weak coupling limit $b\rightarrow 0$ \cite{luk}
\beq
\lambda_{Liouv}(\alpha) \rightarrow \frac{b}{4} \lambda_{KG}(\alpha)
\eeq

\section{Discussion}

In this paper we have discussed the different aspects of the duality between the
calculation of the Wilson polygons and amplitudes in SUSY gauge theories
focusing mainly on the one loop correspondence. It turns out that the duality
for the MHV amplitude can be
explicitly derived in the one-loop case. The derivation is remarkably simple
and involves only the proper change of the variables and the relation between the
Feynman integrals in the different space-time dimensions. The Wilson polygon
to some extend can be thought as placed in the space of the Feynman parameters
and it is in this space the change of variables is important.  It was shown that
the UV behavior of the Wilson polygon precisely maps into  IR behavior of the
amplitude which explains the correspondence between the regularizations observed earlier.

The change of variables found works well for the MHV amplitude only
which can be expressed
in terms of  two mass-easy box diagrams and the  generalization of the
duality for the NMHV cases is required.
Note that we have identified the key feature of the MHV kinematics - only
in this case the integration over reparametrizations is decoupled
which is not true for the rest of the cases. Therefore  one could expect
for the generic kinematics the emergence of the correlators of the Liouville
modes responsible for the reparametrizations of the boundary contour.
We consider the similar duality for the three-point
function with one external particle on-shell. It was shown that the duality can
be formulated upon the  insertion of the peculiar vertex operator into the
Wilson triangle. We consider this example as providing the possible way for the
generalization of the duality for the polarizaton-sensitive case. Let us emphasize
that  SUSY was not essentially used in our one-loop derivation of the duality.
Probably the duality can be similarly developed for the non-SUSY
theories as well.

It is worth to make more general comment concerning the relation
of our analysis
with the moduli space geometry.  In the approach of \cite{gopa,aha}
the Schwinger parameters get mapped generically into the radial
coordinate in $AdS_5$ and the moduli space of the complex structures
$M_{g,n}$ where n is related to the number of the external legs in the
amplitude. That is the Schwinger parametrization is closely related
to the B model. On the other hand in our paper we exploited
the picture with the emergent Kahler moduli which happens in $A$ model.
In principle one could imagine that a kind of the mirror transform
on the level of the Feynman diagrams can be formulated and it
would be very interesting to investigate this issue further.
Note also that the $A$ model under consideration allows the target space
effective description is terms of the effective noncommutative gauge
theory \cite{foam}. We hope to discuss the possible relation
between the Wilson loops in $D=6$ we have discussed with
the corresponding object in the effective target space $D=6$ gauge
theory elsewhere.

The duality implies that a kind of the unitary technique can be developed for
the calculation of the Wilson polygon as well. We have formulated the cut
procedure for the one-loop Wilson polygon which involves the integration
along the cut with the peculiar vertex-like operator. Along this line of reasoning
we have also derived the analogue of the Landau equations for the singularities
for the Wilson polygon in terms of the Feynman parameters. Let us
emphasize that the geometry behind the Landau equations has a lot in common
with the generic hyperbolic geometry behind the one-loop amplitudes.
Actually the generic off-shell box diagram calculates the hyperbolic volume
of the simplex defined by the kinematical invariants that is all divergences
emerging upon some external particle tends to be on-shell are expected
to carry some geometrical information about the initial hyperbolic geometry.
We have shown that at the one-loop level  this happens indeed.

In this paper we have discussed the one loop case only
hence it would be very interesting
to extend this analysis to the higher loops.
The approach to the all-loop answer based on the quantum
geometry of the momentum space
suggested in \cite{gor} could be useful. Another promising development
concerns the relation with the geometry of the knots which
emerges because of the relation with the volumes of the hyperbolic spaces
identified with the knot complements.

We are grateful to G. Korchemsky, N. Nekrasov, Yu. Makeenko and A. Rosly for the
useful discussions. The work was supported in part by grants PICS- 07-0292165(A.G')
and 09-02-00308(A.G., A.Z.).
A.G. thanks FTPI at University of Minnesota where the part of the work
was done for the kind hospitality and support.
A.Z. thanks ICTP at Trieste where the part of the work
was done for the kind hospitality and support.

\section*{Appendix A \qquad Connection of scalar integrals in different dimensions}

Here we briefly explain the connection between the scalar integrals in different dimensions \cite{Nizic2}.
Suppose, we have the following scalar integral
\begin{eqnarray} \label{conn1}
I^{N}(D;{\nu_{k}}) \equiv - i \pi^{-\frac{D}{2}} (\mu^2)^{\epsilon} \int d^{D}l \frac{1}{A^{\nu_{1}}_{1}A^{\nu_{2}}_{2}...A^{\nu_{N}}_{N}} \\ \nonumber
\end{eqnarray}

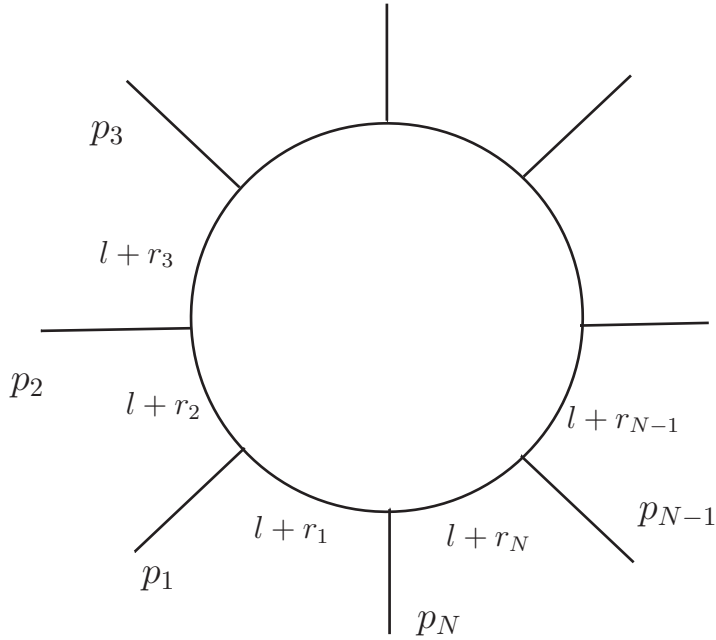
\begin{center}
\fcolorbox{white}{white}{
  \begin{picture}(267,239) (68,-79)
    \SetWidth{1.0}
    \SetColor{Black}
    \Arc(205,41)(73.007,1,361)
    \Line(111,-47)(152,-8)
    \Line(76,36)(132,37)
    \Line(256,94)(296,132)
    \Line(277,38)(325,39)
    \Line(206,-31)(206,-78)
    \Line(205,115)(205,159)
    \Line(255,-11)(297,-51)
    \Line(150,90)(108,130)
    \Text(95,106)[lb]{\large{\Black{$p_{3}$}}}
    \Text(217,-79)[lb]{\large{\Black{$p_{N}$}}}
    \Text(300,-38)[lb]{\large{\Black{$p_{N-1}$}}}
    \Text(228,-47)[lb]{\normalsize{\Black{$l+r_{N}$}}}
    \Text(156,-44)[lb]{\normalsize{\Black{$l+r_{1}$}}}
    \Text(108,3)[lb]{\normalsize{\Black{$l+r_{2}$}}}
    \Text(98,60)[lb]{\normalsize{\Black{$l+r_{3}$}}}
    \Text(273,-3)[lb]{\normalsize{\Black{$l+r_{N-1}$}}}
    \Text(114,-62)[lb]{\large{\Black{$p_{1}$}}}
    \Text(65,11)[lb]{\large{\Black{$p_{2}$}}}
  \end{picture}
}
\end{center}

Than it can be shown that
\begin{eqnarray} \label{conn}
I^{N} (D-2;{\nu_{k}}) = \sum_{i=1}^{N} z_{i} I^{N} (D-2;{\nu_{k} - \delta_{k i}}) \\ \nonumber
+ (D - 1 - \sum_{j=1}^{N} \nu_{j}) z_{0} I^{N} (D;{\nu_{k}})
\end{eqnarray}
where

\begin{eqnarray} \label{conn}
\sum_{i=1}^{N} (r_{i} - r_{j})^2 z_{i} &=& 1 \\ \nonumber
z_{0} &=& \sum_{i=1}^{N} z_{i}
\end{eqnarray}
In the main body of the text we choose $D=6 + 2 \epsilon$, $N=4$, $\nu_{i}=1$.

\section*{Appendix B \qquad Sum over states in terms of dual superconformal invariants}

It is convenient to use ${\cal N}=4$ on-shell formulation of ${\cal N}=4$ SYM, in which all states are encoded in one super-wavefunction
\begin{eqnarray} \label{swave}
  \Phi(p,\eta) &=& G^{+}(p) + \eta^{A} \Gamma_{A}(p) + \frac{1}{2} \eta^{A} \eta^{B} S_{A B}(p) + \frac{1}{3!} \eta^{A} \eta^{B} \eta^{C} \epsilon_{ABCD} \widetilde{\Gamma}^{D}(p) \nonumber\\
  &+& \frac{1}{4!} \eta^{A} \eta^{B} \eta^{C} \eta^{D} \epsilon_{ABCD} G^{-}(p)
\end{eqnarray}
We are interested in cuts of superamplitudes for $n$ particles
\begin{eqnarray} \label{sam}
  {\cal A}_{n}(\lambda,\tilde{\lambda},\eta)={\cal A}_{n}(\Phi_{1},...,\Phi_{n})
\end{eqnarray}
Using of superamplitudes formalism one can easily obtain particular configuration of states using known projectors.
As usually we use the two-component spinor formalism, where
$p^{\alpha \dot{\alpha}}_{i}= \lambda^{\alpha}_{i} \tilde{\lambda}^{\dot{\alpha}}_{i}$ and $\langle i | j \rangle = \lambda^{\alpha}_{i} \lambda_{j \alpha}$.

While the N$^{k}$MHV amplitude can be presented in terms of nested sums which are quite cumbersome expressions for MHV and NMHV cases are pretty simple
\begin{eqnarray} \label{mhv}
  {\cal A}_{n}^{MHV}=\frac{\delta^{(4)}(\sum_{i=1}^{n} \lambda^{\alpha}_{i} \tilde{\lambda}^{\dot{\alpha}}_{i}) \delta^{(8)}(\sum_{i=1}^{n} \lambda^{\alpha}_{i} \eta^{A}_{i})}{\langle1 | 2\rangle...\langle n-1 | n \rangle \langle n | 1 \rangle}
\end{eqnarray}
where the second (Grassmann) delta-function makes the supersymmetry manifest.
\begin{eqnarray} \label{nmhv}
  {\cal A}_{n}^{NMHV}=\frac{\delta^{(4)}(\sum_{i=1}^{n} \lambda^{\alpha}_{i} \tilde{\lambda}^{\dot{\alpha}}_{i}) \delta^{(8)}(\sum_{i=1}^{n} \lambda^{\alpha}_{i} \eta^{A}_{i})}{\langle1 | 2\rangle...\langle n-1 | n \rangle \langle n | 1 \rangle} \sum_{(i,j)} R_{k;ij}
\end{eqnarray}
and all indexes are understood in the following way $ i + n \equiv i $. Then $k+2 \leq i<j \leq n + k-1$ and $ j - i \geq 2 $. $R_{k,ij}$ are dual conformal
invariants which are given by the following expressions

\begin{eqnarray} \label{dci}
  R_{k;ij}=\frac{\langle i | i-1 \rangle \langle j | j-1 \rangle \delta^{(4)} (\Xi_{k;ij})}{x_{i j}^2 \langle k | x_{k i} x_{ij} | j\rangle \langle k| x_{k i} x_{i j} |j-1 \rangle \langle k| x_{k j} x_{j i} |i \rangle \langle k | x_{k j} x_{j i} |i-1 \rangle }
\end{eqnarray}
Here the $\Xi_{k;ij}$ is
\begin{eqnarray} \label{xiDH}
  \Xi_{k;ij} = \langle k| x_{k i} x_{i j} |\theta_{j k} \rangle + \langle k| x_{k j} x_{j i} | \theta_{i k} \rangle
\end{eqnarray}
thus in this language $R$ depends on $n-2$ momenta.

For supermomentum delta-functions we will widely use the following identity ~\cite{Khoze} which could be easily proved
\begin{eqnarray} \label{df1}
   \delta^{(8)} (I) \delta^{(8)} (J) = \delta^{(8)} (I + J) \delta^{(8)} (J)
\end{eqnarray}

The summation over states is equivalent to the integration over $\int d^4 \eta_{cut}$. For particular cuts,
it was done in term of MHV vertex expansion in ~\cite{Freedman}. Here we present some results in simple forms
using the language of dual conformal invariants $R$.

Thus, summation over states in L-loop cut is obtained by

\begin{eqnarray} \label{sum}
   {\cal A}_{cut}^{sum}=\int d^4 \eta_{cut,1} d^4 \eta_{cut,2}d^4 ... \eta_{cut,L}d^4 \eta_{cut,L+1} {\cal A}_{left}^{tree} \ast {\cal A}_{right}^{tree}
\end{eqnarray}

For N$^{L-1}$MHV $\times$ MHV cuts using (\ref{df1}) we can interpret the summation over states as the action of projection operators.

The result for N$^{L-1}$MHV $\times$ MHV cut, thus, reads as
\begin{eqnarray} \label{sum}
   {\cal A}_{cut}^{sum}= \frac{\delta^{(8)}(ext) }{\langle l_{L} | l_{L+1} \rangle^4} A^{gluons,tree}_{N^{L-1} MHV split} (++...+_{ext};--...-_{loop}) A^{gluons,tree}_{MHV, right}(--+...+_{loop};+...+_{ext})
\end{eqnarray}

If we cut MHV diagram at one loop (suppose loop momenta are $l_{1}$ and $l_{2}$) then for state sum we have
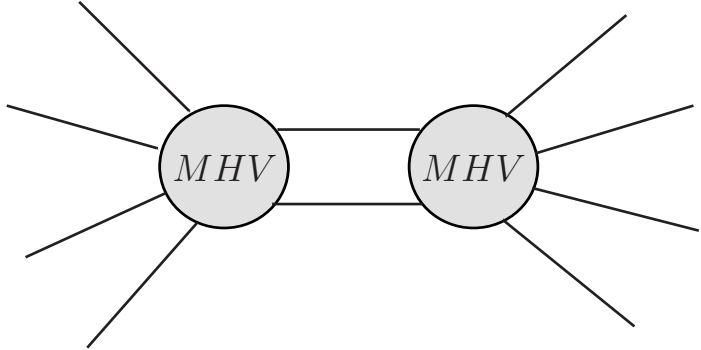
\begin{center}
\fcolorbox{white}{white}{
  \begin{picture}(260,132) (8,-17)
    \SetWidth{1.0}
    \SetColor{Black}
    \GOval(90,52)(23,24)(0){0.882}
    \GOval(183,52)(23,24)(0){0.882}
    \Line(77,73)(36,114)
    \Line(65,59)(9,75)
    \Line(68,42)(16,18)
    \Line(80,31)(39,-16)
    \Line(195,71)(240,109)
    \Line(206,57)(265,75)
    \Line(205,44)(267,28)
    \Line(194,32)(243,-8)
    \Line(110,66)(163,66)
    \Line(108,38)(164,38)
    \Text(71,47)[lb]{\large{\Black{$MHV$}}}
    \Text(164,47)[lb]{\large{\Black{$MHV$}}}
  \end{picture}
}
\end{center}

\begin{eqnarray} \label{sum}
   {\cal A}_{cut}^{sum}= \frac{\delta^{(8)}(ext) }{\langle l_{1} | l_{2} \rangle^4} A^{gluons,tree}_{MHV} (++...+_{ext};--_{loop}) A^{gluons,tree}_{MHV, right}(--_{loop};+...+_{ext})
\end{eqnarray}
which agrees with formulas obtained in the literature.

To our purposes we need divide it on the tree-level amplitude which we have cut. That would be the operator which glues together Wilson loops:

\begin{eqnarray} \label{sum}
  V_{W} &=& \frac{\prod_{ext} \langle i|i+1 \rangle}{\langle l_{1} | l_{2} \rangle^4} A^{gluons,tree}_{MHV} (++...+_{ext};--_{loop}) A^{gluons,tree}_{MHV, right}(--_{loop};+...+_{ext}) \\ \nonumber
  &=&\frac{\langle i | i-1 \rangle \langle j | j-1 \rangle}{\langle l_{1} | j\rangle \langle l_{1} |j-1 \rangle \langle l_{2} |i \rangle \langle l_{2} |i-1 \rangle} <l_{1}|l_{2}>^2
\end{eqnarray}
Another case of special interest for us is anti-MHV $\times$ MHV cut when we obtain
\begin{eqnarray} \label{sum}
   {\cal A}_{cut}^{sum}= \frac{\delta^{(8)}(ext) }{\langle l_{L} | l_{L+1} \rangle^4} A^{gluons,tree}_{\overline{MHV}} (++_{ext};--...-_{loop}) A^{gluons,tree}_{MHV, right}(--+...+_{loop};+...+_{ext})
\end{eqnarray}


\begin{thebibliography}{99}
\bibitem{witten}
  E.~Witten,
  ``Perturbative gauge theory as a string theory in twistor space,''
  Commun.\ Math.\ Phys.\  {\bf 252}, 189 (2004)
  [arXiv:hep-th/0312171].


\bibitem{am}
  L.~F.~Alday and J.~M.~Maldacena,
  ``Gluon scattering amplitudes at strong coupling,''
  JHEP {\bf 0706} (2007) 064
  [arXiv:0705.0303 [hep-th]].

\bibitem{pm}
   A.~M.~Polyakov,
  ``The wall of the cave,''
  Int.\ J.\ Mod.\ Phys.\  A {\bf 14}, 645 (1999)
  [arXiv:hep-th/9809057]. \\
  J.~McGreevy and A.~Sever,
  ``Planar scattering amplitudes from Wilson loops,''
  JHEP {\bf 0808} (2008) 078
  [arXiv:0806.0668 [hep-th]].

\bibitem{kor}
  J.~M.~Drummond, G.~P.~Korchemsky and E.~Sokatchev,
  ``Conformal properties of four-gluon planar amplitudes and Wilson loops,''
  Nucl.\ Phys.\  B {\bf 795}, 385 (2008)
  [arXiv:0707.0243 [hep-th]].\\
   J.~M.~Drummond, J.~Henn, G.~P.~Korchemsky and E.~Sokatchev,
  ``On planar gluon amplitudes/Wilson loops duality,''
  Nucl.\ Phys.\  B {\bf 795}, 52 (2008)
  [arXiv:0709.2368 [hep-th]].
\bibitem{trav}
A.~Brandhuber, P.~Heslop and G.~Travaglini,
  ``MHV Amplitudes in N=4 Super Yang-Mills and Wilson Loops,''
  Nucl.\ Phys.\  B {\bf 794}, 231 (2008)
  [arXiv:0707.1153 [hep-th]].
\bibitem{six1}
  Z.~Bern, L.~J.~Dixon, D.~A.~Kosower, R.~Roiban, M.~Spradlin, C.~Vergu and A.~Volovich,
  ``The Two-Loop Six-Gluon MHV Amplitude in Maximally Supersymmetric Yang-Mills
  Theory,''
  arXiv:0803.1465 [hep-th].
\bibitem{six2}
  J.~M.~Drummond, J.~Henn, G.~P.~Korchemsky and E.~Sokatchev,
 ``Hexagon Wilson loop = six-gluon MHV amplitude,''
  arXiv:0803.1466 [hep-th].
\bibitem{kor2}
  J.~M.~Drummond, J.~Henn, G.~P.~Korchemsky and E.~Sokatchev,
  ``Dual superconformal symmetry of scattering amplitudes in N=4
  super-Yang-Mills theory,''
  arXiv:0807.1095 [hep-th].
\bibitem{kor3}
  J.~M.~Drummond, J.~Henn, G.~P.~Korchemsky and E.~Sokatchev,
  ``Conformal Ward identities for Wilson loops and a test of the duality with
  gluon amplitudes,''
  arXiv:0712.1223 [hep-th].
\bibitem{kom}
  Z.~Komargodski,
  ``On collinear factorization of Wilson loops and MHV amplitudes in N=4 SYM,''
  JHEP {\bf 0805}, 019 (2008)
  [arXiv:0801.3274 [hep-th]].\\
    L.~F.~Alday and J.~Maldacena,
  ``Minimal surfaces in AdS and the eight-gluon scattering amplitude at strong
  coupling,''
  arXiv:0903.4707 [hep-th].


\bibitem{alday}
   L.~F.~Alday and R.~Roiban,
  ``Scattering Amplitudes, Wilson Loops and the String/Gauge Theory
  Correspondence,''
  arXiv:0807.1889 [hep-th].
\bibitem{ber}
  N.~Berkovits and J.~Maldacena,
  ``Fermionic T-Duality, Dual Superconformal Symmetry, and the Amplitude/Wilson Loop Connection,''
  arXiv:0807.3196 [hep-th].\\
N.~Beisert, R.~Ricci, A.~Tseytlin and M.~Wolf,
  ``Dual Superconformal Symmetry from AdS5 x S5 Superstring Integrability,''
  arXiv:0807.3228 [hep-th].
\bibitem{yangian}
  J.~M.~Drummond, J.~M.~Henn and J.~Plefka,
  ``Yangian symmetry of scattering amplitudes in N=4 super Yang-Mills theory,''
  arXiv:0902.2987 [hep-th].
\bibitem{Tarasov}
 O.~V.~Tarasov,
  ``Reduction of Feynman graph amplitudes to a minimal set of basic
  integrals,''
  Acta Phys.\ Polon.\  B {\bf 29}, 2655 (1998)
  [arXiv:hep-ph/9812250].
 \bibitem{Nizic2}
 G.~Duplancic and B.~Nizic,
  ``Reduction method for dimensionally regulated one-loop N-point Feynman
  integrals,''
  Eur.\ Phys.\ J.\  C {\bf 35}, 105 (2004)
  [arXiv:hep-ph/0303184].
 \bibitem{Nizic}
 G.~Duplancic and B.~Nizic,
  ``Dimensionally regulated one-loop box scalar integrals with massless
  internal lines,''
  Eur.\ Phys.\ J.\  C {\bf 20}, 357 (2001)
  [arXiv:hep-ph/0006249].




\bibitem{gopa}
  R.~Gopakumar,
  ``From free fields to AdS,''
  Phys.\ Rev.\  D {\bf 70}, 025009 (2004)
  [arXiv:hep-th/0308184].

\bibitem{aha}
  O.~Aharony, Z.~Komargodski and S.~S.~Razamat,
  ``On the worldsheet theories of strings dual to free large N gauge
  theories,''
  JHEP {\bf 0605}, 016 (2006)
  [arXiv:hep-th/0602226].\\
 O.~Aharony, J.~R.~David, R.~Gopakumar, Z.~Komargodski and S.~S.~Razamat,
  ``Comments on worldsheet theories dual to free large N gauge theories,''
  Phys.\ Rev.\  D {\bf 75}, 106006 (2007)
  [arXiv:hep-th/0703141].\\
R.~Gopakumar,
  ``From free fields to AdS. III,''
  Phys.\ Rev.\  D {\bf 72}, 066008 (2005)
  [arXiv:hep-th/0504229].
\bibitem{gorly}
  A.~Gorsky and V.~Lysov,
  ``From effective actions to the background geometry,''
  Nucl.\ Phys.\  B {\bf 718}, 293 (2005)
  [arXiv:hep-th/0411063].
\bibitem{dav}
  A.~I.~Davydychev and R.~Delbourgo,
  ``A geometrical angle on Feynman integrals,''
  J.\ Math.\ Phys.\  {\bf 39}, 4299 (1998)
  [arXiv:hep-th/9709216].

\bibitem{tausk}
  A.~I.~Davydychev and J.~B.~Tausk,
  ``A Magic connection between massive and massless diagrams,''
  Phys.\ Rev.\  D {\bf 53}, 7381 (1996)
  [arXiv:hep-ph/9504431].
\bibitem{dav2}
  N.~I.~Usyukina and A.~I.~Davydychev,
  ``New results for two loop off-shell three point diagrams,''
  Phys.\ Lett.\  B {\bf 332}, 159 (1994)
  [arXiv:hep-ph/9402223].
\bibitem{bds}
Z.~Bern, L.~J.~Dixon and V.~A.~Smirnov,
  ``Iteration of planar amplitudes in maximally supersymmetric Yang-Mills
  theory at three loops and beyond,''
  Phys.\ Rev.\  D {\bf 72}, 085001 (2005)
  [arXiv:hep-th/0505205].

\bibitem{gor}
A.~ Gorsky, "Quantum geometry of the momentum space and  amplitudes in the N=4
SYM theory", In Proceedings of "Quarks-2008*"(May 2008) and memorial "Landau-100"
(June 2008) Conferences \\
A.~ Gorsky, To appear



\bibitem{bgk}
  A.~V.~Belitsky, A.~S.~Gorsky and G.~P.~Korchemsky,
  ``Gauge / string duality for QCD conformal operators,''
  Nucl.\ Phys.\  B {\bf 667}, 3 (2003)
  [arXiv:hep-th/0304028].


\bibitem{gorsky}
  A.~Gorsky,
  ``Spin chains and gauge / string duality,''
  Theor.\ Math.\ Phys.\  {\bf 142}, 153 (2005)
  [Teor.\ Mat.\ Fiz.\  {\bf 142}, 179 (2005)]
  [arXiv:hep-th/0308182].


\bibitem{brod}
  D.~J.~Broadhurst,
  ``Solving differential equations for 3-loop diagrams: Relation to  hyperbolic
  geometry and knot theory,''
  arXiv:hep-th/9806174.


\bibitem{Sok1} J.Drummond, J.Henn, G.Korchemsky, E.Sokatchev, [hep-th/0808.0491]; \\
 J.~M.~Drummond, J.~Henn, G.~P.~Korchemsky and E.~Sokatchev,
  ``Generalized unitarity for N=4 super-amplitudes,''
  arXiv:0808.0491 [hep-th].
\bibitem{Tree}
 J.~M.~Drummond and J.~M.~Henn,
  ``All tree-level amplitudes in N=4 SYM,''
  arXiv:0808.2475 [hep-th].
\bibitem{Freedman}
 H.~Elvang, D.~Z.~Freedman and M.~Kiermaier,
  ``Recursion Relations, Generating Functions, and Unitarity Sums in N=4 SYM
  Theory,''
  arXiv:0808.1720 [hep-th].
\bibitem{Khoze}
 G.~Georgiou, E.~W.~N.~Glover and V.~V.~Khoze,
  ``Non-MHV Tree Amplitudes in Gauge Theory,''
  JHEP {\bf 0407}, 048 (2004)
  [arXiv:hep-th/0407027].

\bibitem{polyakov}
  A.~M.~Polyakov,
  ``String theory and quark confinement,''
  Nucl.\ Phys.\ Proc.\ Suppl.\  {\bf 68}, 1 (1998)
  [arXiv:hep-th/9711002].\\
  Y.~Makeenko and P.~Olesen,
  ``Implementation of the Duality between Wilson loops and Scattering
  Amplitudes in QCD,''
  Phys.\ Rev.\ Lett.\  {\bf 102}, 071602 (2009)
  [arXiv:0810.4778 [hep-th]].
\bibitem{Landau}
  L.~D.~Landau,
  ``On analytic properties of vertex parts in quantum field theory,''
  Nucl.\ Phys.\  {\bf 13}, 181 (1959).
\bibitem{Cutkosky}
  R.~E.~Cutkosky,
  ``Singularities and discontinuities of Feynman amplitudes,''
  J.\ Math.\ Phys.\  {\bf 1} (1960) 429.
\bibitem{Lowe}
  R.~Eden, P.~Landshoff, D.~Olive, J.~Polkinghorne,
  ``The analytic S-matrix,'' (1966);
\bibitem{Smirnov}
  V.~Smirnov,
  ``Evaluating Feynman integrals,''
    Tracts Mod.Phys.211:1-244, (2004)
\bibitem{bernNMHV}
 Z.~Bern, L.~J.~Dixon and D.~A.~Kosower,
  ``All next-to-maximally helicity-violating one-loop gluon amplitudes in ${\cal N}=4$ super-Yang-Mills theory,''
  Phys.\ Rev.\  D {\bf 72}, 045014 (2005)
  arXiv:0412210 [hep-th].
\bibitem{DavydRed} A.Davydychev,
  A.~I.~Davydychev,
  ``A Simple formula for reducing Feynman diagrams to scalar integrals,''
  Phys.\ Lett.\  B {\bf 263}, 107 (1991).
\bibitem{Brand2}
 C.~Anastasiou, A.~Brandhuber, P.~Heslop, V.~Khoze, B.~Spence, G.~Travaglini,
 ``Two-Loop Polygon Wilson Loops in N=4 SYM,''
  arXiv:0902.2245 [hep-th].
\bibitem{hikami}
  K.~Hikami,
  ``Generalized Volume Conjecture and the A-Polynomials: The Neumann-Zagier
  Potential Function as a Classical Limit of Quantum Invariant,''
  J.\ Geom.\ Phys.\  {\bf 57}, 1895 (2007)
  [arXiv:math/0604094].\\
   T.~Dimofte, S.~Gukov, J.~Lenells and D.~Zagier,
  ``Exact Results for Perturbative Chern-Simons Theory with Complex Gauge
  Group,''
  arXiv:0903.2472 [hep-th].

\bibitem{foam}
  A.~Iqbal, N.~Nekrasov, A.~Okounkov and C.~Vafa,
  ``Quantum foam and topological strings,''
  JHEP {\bf 0804}, 011 (2008)
  [arXiv:hep-th/0312022].
  \bibitem{luk}
  S.~L.~Lukyanov,
  ``Correlators of the Jost functions in the Sine-Gordon model,''
  Phys.\ Lett.\  B {\bf 325}, 409 (1994)
  [arXiv:hep-th/9311189].


\end{thebibliography}
\end{document}